\documentclass[pdflatex,sn-mathphys]{sn-jnl}
\usepackage{siunitx}
\usepackage[T1]{fontenc}
\jyear{2021}%

\begin{document}

\title[Steering self-organisation through confinement]{Steering self-organisation through confinement}

\author*[1,2]{\fnm{Nuno A. M.} \sur{Araújo}}\email{nmaraujo@fc.ul.pt}

\author*[3,4]{\fnm{Liesbeth M. C.} \sur{Janssen}}\email{l.m.c.janssen@tue.nl}

\author[5]{\fnm{Thomas} \sur{Barois}}

\author[6]{\fnm{Guido} \sur{Boffetta}}

\author[7,8]{\fnm{Itai} \sur{Cohen}}

\author[3]{\fnm{Alessandro} \sur{Corbetta}}

\author[9]{\fnm{Olivier} \sur{Dauchot}}

\author[10]{\fnm{Marjolein} \sur{Dijkstra}}

\author[11]{\fnm{William M.} \sur{Durham}\textsuperscript{11}}

\author[12]{\fnm{Audrey} \sur{Dussutour}}

\author[13]{\fnm{Simon} \sur{Garnier}}

\author[4,14]{\fnm{Hanneke} \sur{Gelderblom}}
 
\author[15,16]{\fnm{Ramin} \sur{Golestanian}}

\author[17]{\fnm{Lucio} \sur{Isa}}

\author[18]{\fnm{Gijsje H.} \sur{Koenderink}}

\author[19]{\fnm{Hartmut} \sur{L\"owen}}

\author[20]{\fnm{Ralf} \sur{Metzler}}

\author[21,22]{\fnm{Marco} \sur{Polin}}

\author[9]{\fnm{C. Patrick} \sur{Royall}}

\author[23]{\fnm{An{\dj}ela} \sur{Šarić}}

\author[24]{\fnm{Anupam} \sur{Sengupta}}

\author[25]{\fnm{Cécile} \sur{Sykes}}

\author[26]{\fnm{Vito} \sur{Trianni}}

\author[21]{\fnm{Idan} \sur{Tuval}}

\author[27]{\fnm{Nicolas} \sur{Vogel}}

\author[16]{\fnm{Julia M.} \sur{Yeomans}}

\author[28]{\fnm{Iker} \sur{Zuriguel}}

\author*[29]{\fnm{Alvaro} \sur{Marin}} \email{a.marin@utwente.nl}

\author*[30]{\fnm{Giorgio} \sur{Volpe}} \email{g.volpe@ucl.ac.uk}

\affil[1]{\orgdiv{Departamento de Física, Faculdade de Ciências}, \orgname{Universidade de Lisboa}, \orgaddress{\postcode{1749-016} \city{Lisboa}, \country{Portugal}}}

\affil[2]{\orgdiv{Centro de Física Teórica e Computacional, Faculdade de Ciências}, \orgname{Universidade de Lisboa}, \orgaddress{\postcode{1749-016} \city{Lisboa}, \country{Portugal}}}

\affil[3]{\orgdiv{Department of Applied Physics}, \orgname{Eindhoven University of Technology}, \orgaddress{\street{P.O. Box 513}, \postcode{5600 MB} \city{Eindhoven}, \country{The Netherlands}}}

\affil[4]{\orgdiv{Institute for Complex Molecular Systems}, \orgname{Eindhoven University of Technology}, \orgaddress{\street{P.O. Box 513}, \postcode{5600 MB} \city{Eindhoven}, \country{The Netherlands}}}

\affil[5]{\orgname{University of Bordeaux, CNRS, LOMA, UMR 5798}, \orgaddress{\postcode{F-33400} \city{Talence}, \country{France}}}

\affil[6]{\orgdiv{Department of Physics and INFN}, \orgname{University of Torino}, \orgaddress{\street{via Pietro Giuria 1}, \postcode{10125} \city{Torino}, \country{Italy}}}

\affil[7]{\orgdiv{Kavli Institute at Cornell for Nanoscale Science}, \orgname{Cornell University}, \orgaddress{\city{Ithaca}, \state{New York}, \country{USA}}}

\affil[8]{\orgdiv{Laboratory of Atomic and Solid-State Physics}, \orgname{Cornell University}, \orgaddress{\city{Ithaca}, \state{New York}, \country{USA}}}

\affil[9]{\orgdiv{Gulliver UMR CNRS 7083}, \orgname{ESPCI Paris, Universit\'{e} PSL}, \orgaddress{\postcode{75005} \city{Paris}, \country{France}}}

\affil[10]{\orgdiv{Soft condensed matter, Department of Physics, Debye institute for Nanomaterials Science}, \orgname{Utrecht University}, \orgaddress{\street{Princetonplein 1}, \postcode{3584 CC} \city{Utrecht}, \country{The Netherlands}}}

\affil[11]{\orgdiv{Department of Physics and Astronomy}, \orgname{University of Sheffield}, \orgaddress{\street{Hounsfield Road}, \city{Sheffield} \postcode{S3 7RH}, \country{United Kingdom}}}

\affil[12]{\orgdiv{Research Centre on Animal Cognition (CRCA), Centre for Integrative Biology (CBI)}, \orgname{Toulouse University, CNRS, UPS}, \orgaddress{\city{Toulouse}, \postcode{31062}, \state{AD}, \country{France}}}

\affil[13]{\orgdiv{Department of Biological Sciences}, \orgname{New Jersey Institute of Technology}, \orgaddress{\city{Newark} \state{NJ} \postcode{07102}, \country{USA}}}

\affil[14]{\orgdiv{Department of Applied Physics and J.M. Burgers Center for Fluid Dynamics}, \orgname{Eindhoven University of Technology}, \orgaddress{\street{P.O. Box 513}, \postcode{5600 MB} \city{Eindhoven}, \country{The Netherlands}}}

\affil[15]{\orgdiv{Max Planck Institute for Dynamics and Self-Organization (MPI-DS)}, \orgaddress{\postcode{37077}, \city{G\"ottingen}, \country{Germany}}}

\affil[16]{\orgdiv{Rudolf Peierls Centre for Theoretical Physics}, \orgname{University of Oxford}, \orgaddress{\city{Oxford} \postcode{OX1 3PU}, \country{United Kingdom}}}

\affil[17]{\orgdiv{Laboratory for Soft Materials and Interfaces, Department of Materials}, \orgname{ETH Z\"{u}rich}, \orgaddress{\postcode{8093} \city{Z\"{u}rich}, \country{Switzerland}}}

\affil[18]{\orgdiv{Department of Bionanoscience, Kavli Institute of Nanoscience}, \orgname{Delft University of Technology}, \orgaddress{\postcode{2629 HZ} \city{Delft}, \country{Netherlands}}}

\affil[19]{\orgdiv{Institut f\"ur Theoretische Physik II: Weiche Materie}, \orgname{Heinrich-Heine-Universit\"at D\"usseldorf}, \orgaddress{\street{Universit\"atsstrasse 1}, \postcode{40225}, \city{D\"usseldorf}, \country{Germany}}}

\affil[20]{\orgdiv{Institute of Physics \& Astronomy}, \orgname{University of Potsdam}, \orgaddress{\street{Karl-Liebknecht-Str 24/25}, \postcode{D-14476}, \city{Potsdam-Golm}, \country{Germany}}}

\affil[21]{\orgdiv{Mediterranean Institute for Advanced Studies}, \orgname{IMEDEA UIB-CSIC}, \orgaddress{\street{C/ Miquel Marqués 21}, \postcode{07190} \city{Esporles}, \country{Spain}}}

\affil[22]{\orgdiv{Department of Physics}, \orgname{University of Warwick}, \orgaddress{\street{Gibbet Hill road}, \postcode{CV4 7AL} \city{ Coventry}, \country{United Kingdom}}}

\affil[23]{\orgname{Institute of Science and Technology Austria}, \orgaddress{ \postcode{3400}, \city{Klosterneuburg}, \country{Austria}}}

\affil[24]{\orgdiv{Physics of Living Matter,
Department of Physics and Materials Science}, \orgname{University of Luxembourg}, \orgaddress{\street{162 A, Avenue de la Fa\"iencerie}, \postcode{L-1511}, \country{Luxembourg}}}

\affil[25]{\orgdiv{Laboratoire de Physique de l’École normale supérieure}, \orgname{ENS, Université PSL, CNRS, Sorbonne Université, Université Paris Cité}, \orgaddress{\postcode{F-75005} \city{Paris}, \country{France}}}

\affil[26]{\orgdiv{Institute of Cognitive Sciences and Technologies}, \orgname{CNR}, \orgaddress{\street{Via San Martino della Battaglia 44}, \postcode{00185} \city{Rome}, \country{Italy}}}

\affil[27]{\orgdiv{Institute of Particle Technology}, \orgname{Friedrich-Alexander Universit\"at Erlangen-N\"urnberg}, \orgaddress{\street{Cauerstrasse 4}, \postcode{91058}, \city{Erlangen}, \country{Germany}}}

\affil[28]{\orgdiv{Departamento de Física y Matemática Aplicada, Facultad de Ciencias}, \orgname{Universidad de Navarra}, \orgaddress{\city{Pamplona}, \country{Spain}}}

\affil[29]{\orgdiv{Physics of Fluids Group, Mesa+ Institute, Max Planck Center for Complex Fluid Dynamics and J. M. Burgers Center for Fluid Dynamics}, \orgname{University of Twente}, \orgaddress{\postcode{7500AE}, \city{Enschede}, \country{The Netherlands}}}

\affil[30]{\orgdiv{Department of Chemistry}, \orgname{University College London}, \orgaddress{\street{20 Gordon Street}, \city{London} \postcode{WC1H 0AJ}, \country{United Kingdom}}}

\abstract{Self-organisation is the spontaneous emergence of spatio-temporal structures and patterns from the interaction of smaller individual units. Examples are found across many scales in very different systems and scientific disciplines, from physics, materials science and robotics to biology, geophysics and astronomy. Recent research has highlighted how self-organisation can be both mediated and controlled by confinement. Confinement occurs through interactions with boundaries, and can function as either a catalyst or inhibitor of self-organisation. It can then become a means to actively steer the emergence or suppression of collective phenomena in space and time. 
Here, to provide a common framework for future research, we examine the role of confinement in self-organisation and identify overarching scientific challenges across disciplines that need to be addressed to harness its full scientific and technological potential. This framework will not only accelerate the generation of a common deeper understanding of self-organisation but also trigger the development of innovative strategies to steer it through confinement, with impact, e.g., on the design of smarter materials, tissue engineering for biomedicine and crowd management.}

\maketitle

\subsection*{Main}\label{intro}

From molecular aggregates \cite{Whitesides91} to groups of animals \cite{moussaid2009collective} and human crowds \cite{Moussaid11,echeverria2022vortex}, from microswimmers \cite{RevModPhys.88.045006} to granular materials \cite{silbert2002boundary} and robotic swarms \cite{oh2017bio},
examples of systems that self-organise can be found across a wide diversity of length and time scales \cite{Whitesides2002,camazine2020self}. The concept of \emph{self-organisation}, as a modern field of study, came to the fore in the 20$^{th}$ century \cite{prigogine1945etude} 
and defines the spontaneous emergence of large-scale collective structures and patterns in space and/or time from the interaction of many 
individual units \cite{Whitesides2002,camazine2020self}, such as molecules, colloidal particles, cells, animals, robots, pedestrians or even astronomical objects. These units can be highly heterogeneous in size, shape, composition and function (as is often the case in biological systems) or largely identical (as in typical synthetic systems). The units can also be active (e.g.\ molecular motors, cells, animals and pedestrians) or passive (e.g.\ colloids, granular matter and planets), depending on whether they can or cannot transform available energy to perform work.

\begin{figure}[h]%
\centering
\includegraphics[width=1\textwidth]{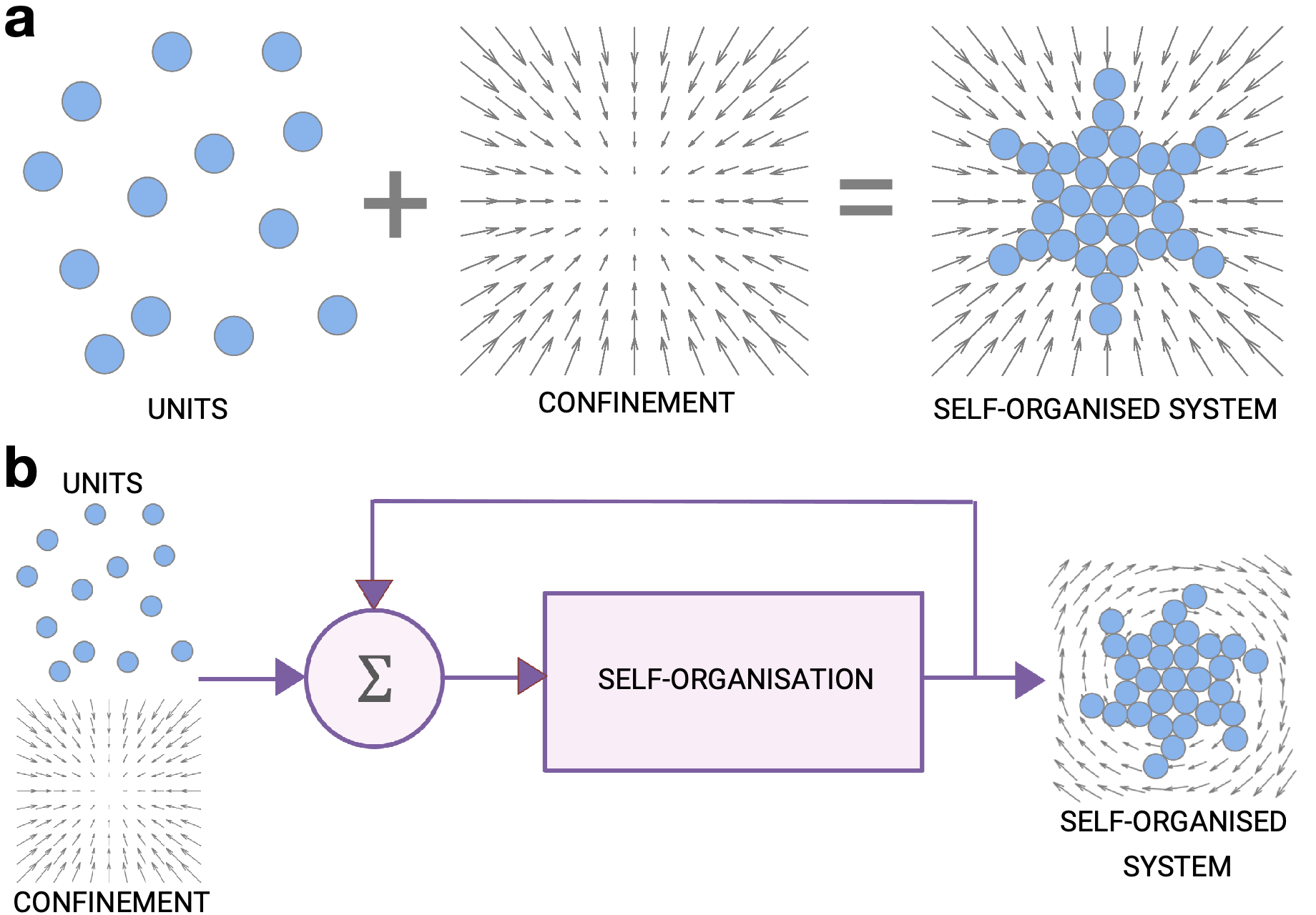}
\caption{{\bf Emergence of structure from confined self-organising units.} ({\bf a}) Self-organisation is the emergence of large-scale structures and patterns from individual units. Confinement can act as a catalyst (as in the diagram) or inhibitor for a self-organising system. The arrows represent an external force field acting on the units. ({\bf b}) Steering self-organisation through confinement requires encoding feedback loops in the process so that units and/or confining elements (inputs) can adapt and evolve with the self-organising system (output). In the schematics, this is visualised by the external force field (confinement) changing to include a curl in response to a control signal (feedback loop) added to the inputs ($\sum$). As a consequence the emerging self-organisation pattern is different from that in {\bf a}.}\label{fig1}
\end{figure}

There are two key features of self-organisation that deserve to be highlighted: first, the self-organised structures extend over much larger length scales than the size of the individual units; second, these structures yield emergent properties and functions, beyond what is achievable by their constituent units alone. This emergence of non-trivial, non-additive collective features on large scales is what makes the topic of self-organisation fascinating. On the one hand, it captures how complex behaviour can develop and evolve from simple units, e.g.\ life itself emerges from a cocktail of lifeless molecules \cite{karsenti2008self}. On the other hand, it provides inspiration to materials scientists and system engineers, who aim to mimic this spontaneous complexity to revolutionise man-made materials and devices \cite{grzybowski2017dynamic}. 

\begin{figure}[h]
\centering
\includegraphics[width=\textwidth]{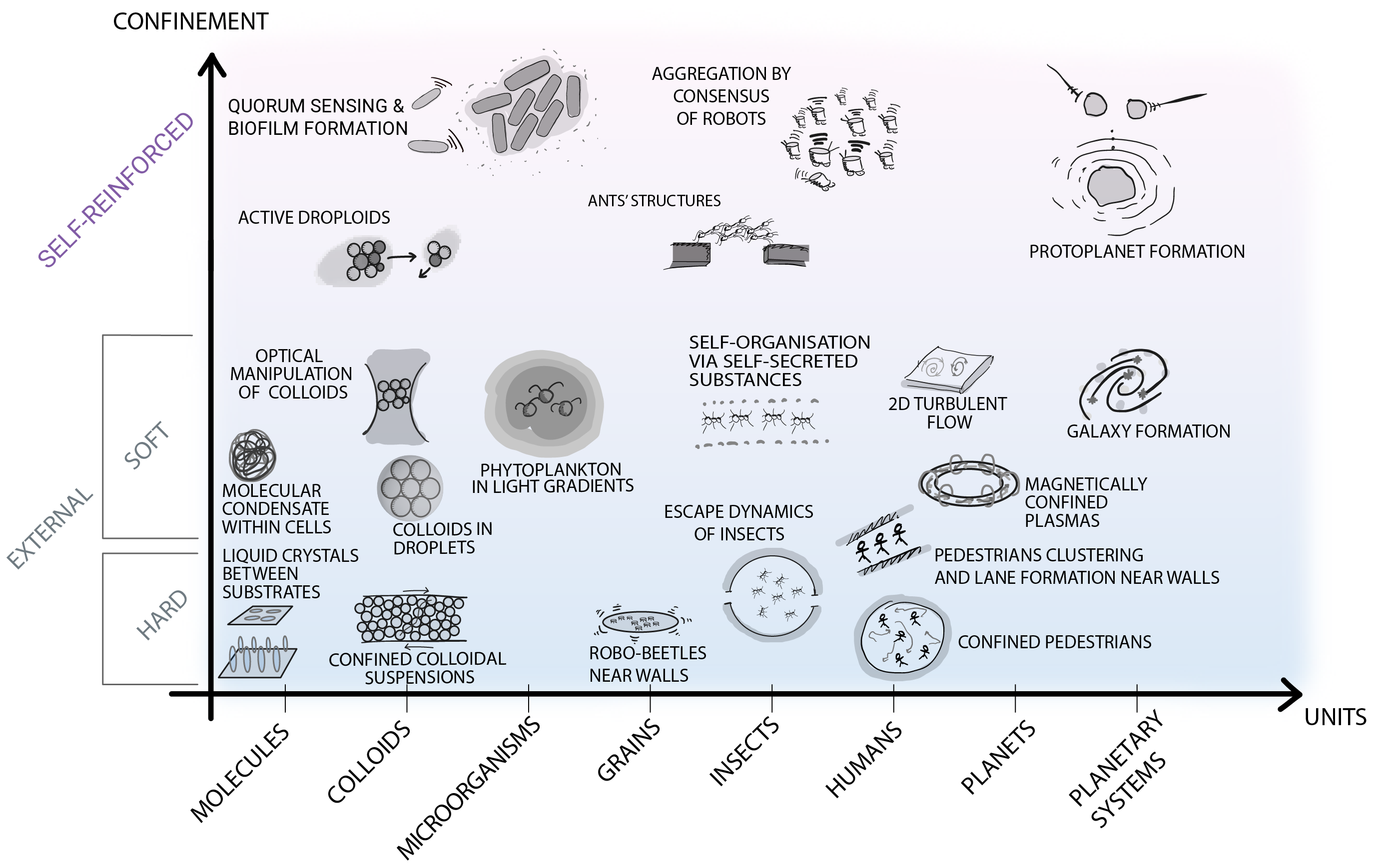}
\caption{{\bf Self-organisation at various length scales under different types of confinement.} The diagram contains selected examples of self-organisation under different types of confinement occurring at different spatial (and time) scales in both natural and man-made systems. The horizontal axis represents the length scale of the self-organising units, from molecular up to astronomical scales. The vertical axis represents the type of confinement ordered based on its complexity and lack of current understanding. At the bottom of the diagram, simpler and better understood forms of confinement are highlighted with blue shading. These include external boundaries (e.g. walls, interfaces and potentials). On the top of the diagram, a different type of confinement is purposely separated from the rest, as less understood but more promising in order to steer self-organisation. These forms of confinement include feedback loops between the self-organising units and the confining features (e.g. in the quorum sensing that precedes microbial biofilm formation \cite{miller2001quorum}, in the information exchange among ants to generate structures such as bridges \cite{mccreery2022hysteresis}, or in the self-induced gravitational attraction in the formation of protoplanets \cite{astroselforganization2018}). 
}\label{fig2}
\end{figure}

In recent years, it has become increasingly evident that \emph{confinement} of the units can influence and even steer the self-organisation process (Fig. \ref{fig1}).
Such confinement may stem, e.g., from the presence of surfaces, interfaces, fields, potentials, flow, and even, in animal and crowd dynamics, less tangible psychological reasons \cite{sieben2017collective}.  
We can thus define confinement in self-organisation as virtually anything 
which causes units to localise to a particular region of space at a given time, leading either to emergence or suppression of collective phenomena. The variety of self-organising systems influenced by confinement is indeed immense, spanning a very wide range of length scales (Fig. \ref{fig2}): from active filaments driven by microscopic molecular motors enclosed within living cells \cite{opathalage2019self}, to the emergence of macroscopic coherent flow structures confined by Earth's atmosphere \cite{boffetta2012two},   
to the formation of entire galaxies under the pull of the gravitational potentials of black holes \cite{kravtsov2012formation}. While confinement is not always required for a system to self-organise \cite{vicsek2012collective}, it can play a pivotal role as either a catalyst or inhibitor for self-organisation. In this regard, one of the most promising applications of confinement in self-organisation is to employ it as a control knob at the hand of researchers and engineers to tune the emergence of collective phenomena. For example, applications of this principle can already be found in the design of molecular sensors on surfaces \cite{taylor2017single}, of scaffolds for tissue engineering~\cite{Neto2019}, and of crowd management strategies via the use of physical barriers \cite{feliciani2020systematic}.

Here, we argue that confinement can be used as a tool to actively steer self-organisation by harnessing non-trivial feedback strategies between a self-organising system and its confining features. This will help generate a deeper understanding of self-organisation in biological systems as well as guide the development of innovative strategies to control self-organisation in man-made ones. To achieve this goal, a concerted effort across disciplines is needed. So far, efforts to understand and control self-organisation under confinement have indeed focused primarily on specific systems with their own particular length and time scales. However, there are many analogous questions and technical challenges found across multiple scales, systems and disciplines, which need to be addressed systematically before the full potential of confinement is harnessed to steer self-organisation. To propose a common roadmap towards this shared aim, this article first defines a unified language to discuss the role of confinement in self-organisation and its control. We then identify the most relevant scientific challenges 
and list the conceptual and technological advances required to tackle them. 

\subsection*{The role of confinement in self-organisation}\label{definitions}

Let us first discuss more in depth how confinement influences self-organisation.
For systems that self-organise without confinement (e.g.\ surfactant molecules forming vesicles in solution \cite{doi2013soft}), its presence can either disrupt the phenomenon altogether and/or lead to the formation of alternative structures and patterns (e.g. monolayers of surfactants on surfaces \cite{doi2013soft}). Nonetheless, for other systems that do not self-organise in its absence (e.g. most of the organelles of a living cell \cite{mccusker2020cellular}), confinement is a prerequisite for self-organisation to emerge (e.g. through the cell membrane, which compartmentalises its interior and separates it
from the external world \cite{sezgin2017mystery}).

More specifically, confinement can affect and steer self-organisation in a number of ways: it can directly influence the translational and rotational degrees of freedom of the units; it can alter the nature and strength of the interactions among them and/or introduce new interactions; it can limit the number and type of units that can interact with each other; it can change the phase space of the self-organising system and its underlying energy landscape; it can modify the encounter rates between units and the probability for sequential or parallel reactions to take place; finally, it can also enable cross-talks across several scales. 

We can introduce several non-mutually exclusive definitions to classify confinement depending on its origin and nature: 

\begin{itemize}

    \item \emph{Physical vs.\ effective}. Physical confinement is imposed by the presence of a physical obstacle that confines the phase space available to a system (e.g.\ a flexible membrane for cells \cite{bashirzadeh2019encapsulation} or a rigid wall for microswimmers \cite{Bechinger2019}, robots \cite{scholz2018rotating} or pedestrians \cite{echeverria2022vortex}); in contrast, effective confinement  stems from an apparent or virtual boundary that is mediated by an intrinsic capability of the units to sense or perceive it (e.g.\ the extent of a chemical trail for bacteria \cite{mukherjee2019bacterial} or ants \cite{czaczkes2015trail}, a force field for colloidal particles \cite{Ebert2009}, a time-dependent distribution of resources consumed by microswimmers \cite{Bechinger2019}, the communication range for animals and robots \cite{rubenstein2014programmable}, the gravitational field confining Earth's atmosphere for turbulent flows \cite{boffetta2012two} or the gravitational field of black holes that keeps a galaxy together \cite{kravtsov2012formation}). 
    
    \item \emph{Hard vs.\ soft}. Hard boundaries are not affected by the dynamics of the self-organising system (as in the case of a solid wall \cite{briand2018spontaneously}), while soft boundaries can deform, reshape, adapt and evolve in response to the dynamics of the self-organisation process (as in the case of a flexible cellular membrane \cite{tsai2015shape} or of a fluid interface \cite{bradley2017janus}), hence there is a feedback mechanism between the units and the confining boundary.
    
    \item \emph{Static vs.\ dynamic}. Static confinement is invariant in time (e.g.\ the walls of a microfluidic chamber for microswimmers \cite{sharan2021microfluidics} or the plates used to confine active granular matter \cite{briand2018spontaneously}); dynamic confinement instead varies in time (e.g.
    time-varying chemical gradients acting as confining fields for groups of cells in tissue \cite{kerszberg2007specifying} 
    or cues leading to history-dependent formations for social animals, as in the case of ants following paths previously made by their peers \cite{czaczkes2015trail}).
    
    \item \emph{Positively vs.\ negatively reinforcing}. 
    Positive and negative reinforcement designate situations where the self-organisation process is enhanced (e.g.\ by autoinducers in microbial quorum sensing \cite{mukherjee2019bacterial} or by chemical gradients in tissue formation and proliferation \cite{michailidi2021morphogen}) or disrupted by the presence of confinement (e.g.\ in the reduction of order in crystal formation due to a porous medium \cite{alba2006effects}).  
    
    \item \emph{External vs.\ self-reinforced}. Finally, confinement is often identified as an external feature, i.e. not belonging to the self-organising system. 
    However, in certain fields (e.g.\ in the study of active colloids, social animals, and in swarm robotics), the concept of self-imposed confinement is well accepted to describe situations where the boundaries originate from within the collective dynamics 
    through internal feedback (e.g.\ 
    perceptual cues for lane formation in social animals, such as ants \cite{czaczkes2015trail}). This is illustrated in the top part of Fig. \ref{fig2} and facilitates a completely different type of confinement.
\end{itemize}

Gaining control over self-organisation through confinement requires the scientific community to leverage the more complex forms of confinement mentioned above, taking advantage of effective, soft, dynamic, and self-reinforced boundaries to create externally or internally imposed feedback mechanisms to steer 
the emergence or suppression of collective behaviours in a self-organising system (Fig. \ref{fig2}). 

\begin{figure}[h]%
\centering
\includegraphics[width=0.9\textwidth]{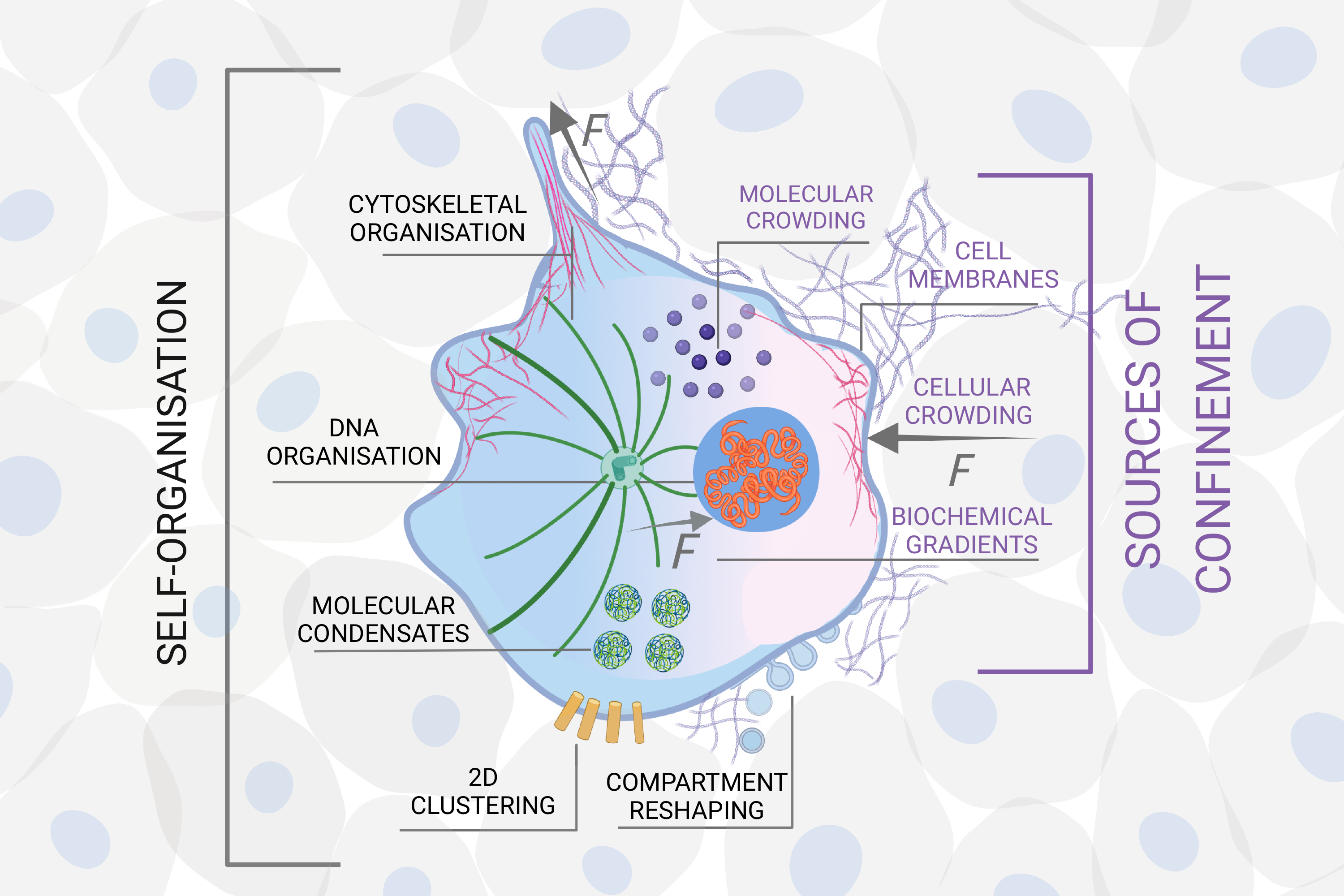}
\caption{{\bf Example of hierarchical self-organisation under confinement in biology.} Hierarchical organisation from molecules to tissue via the formation of macromolecules, cellular organelles and cells. At each stage, self-organised structures become units for further self-organisation subject to a different type of confinement, here illustrated at the molecular, cell and tissue scale. Sources of confinement include, e.g., physical boundaries, mechanical forces ($F$) and chemical gradients. The emergence of complex functionality in biological systems relies on the existence of such hierarchical structures.}\label{fig3}
\end{figure}

\subsection*{Overarching scientific challenges}\label{challenges}

Developing the tools to steer self-organisation through confinement requires us first to gain a deeper fundamental understanding across the systems, scales, and disciplines of how confinement promotes or suppresses the emergence of collective patterns in space and time. We have identified five synergistic areas where further knowledge is required to drive the field forward: universality, heterogeneity, hierarchy, reciprocity, and design by confinement. Whilst universality and heterogeneity are challenges shared with self-organisation in general, the focus here is on the role of confinement.

\begin{itemize}

    \item  \textbf{Universality} aims at understanding whether the patterns observed in a system can be generalised to other systems, scales and disciplines.
    Intrinsically, confinement introduces characteristic (length and/or time) scales to the process, thus potentially jeopardising universality across scales. 
    Nonetheless, establishing the conditions under which system-specific observations can be generalised to other systems and disciplines is pivotal to develop controllable models to study how to steer self-organisation via confinement. 
    
    \item \textbf{Heterogeneity} addresses how variability in the units (e.g. in morphogenesis \cite{teague2016synthetic}, cell differentiation and cancer cells \cite{baron2019unravelling} or in polydisperse colloids \cite{abraham2008energy}) or in the confining element (e.g. heterogeneity in both flow and the distribution of chemicals induced by a porous material \cite{de2021chemotaxis}) influences the emergence of collective behaviour. In particular, the heterogeneity of the confinement in space and time can be employed to influence the self-organising units (e.g. by promoting their segregation or mixing \cite{chaoticadvection2017}) and, vice versa, confinement can be used to trigger the emergence of heterogeneity in the self-organising system (e.g. promoting cell differentiation \cite{baron2019unravelling}). 
    
    \item \textbf{Hierarchy}: 
    Self-organisation can develop hierarchically, when 
    the confinement at a certain scale defines the units at a larger length scale (Fig. \ref{fig3}). For example, in biological systems, molecules (units) self-organise inside a cell confined by its membrane \cite{merino2021self}. The cells themselves can become the units when they form tissues and organs confined, e.g., by the extracellular matrix \cite{chaudhuri2020effects}. 
    Tissues and organs define living entities which can go on to form flocks, herds, schools confined, e.g., by feedback from their senses and perception \cite{ballerini2008interaction}. These groups of animals can then form entire ecosystems confined by their local geography distribution \cite{gude2020bacterial}. 
    In these hierarchical structures, the confining elements at different scales mediate bidirectional interactions and flow of information from smaller to larger scales, and vice versa.  For instance, in biology, the cell membrane is the key confining entity for intracellular self-organisation, but at the same time it defines the cell as an individual unit for multicellular organisation of tissues and organs, thus enabling complex functionalities to emerge. Importantly, the shape and chemical composition of the cell membrane is continuously evolving due to both mechanical and chemical stimuli from the surrounding tissue \cite{balasubramaniam2022active} and from the cell's interior, thus acting as a mediator of the feedback between different scales. 
    The overarching key challenge here is to elucidate, measure and model how (and when) confinement at different scales mediates or separates the cross-talk and interdependence between scales. 
    
    \item \textbf{Reciprocity} can be defined as the formation of dynamic feedback loops between units and confinement, leading to adaptation, responsiveness and even evolution of a self-organising system in response to changing environmental conditions. 
    An example is provided by cell-matrix interactions in wound healing and tissue regeneration~\cite{Foolen2018}, where the extracellular matrix confines cells, forcing them to adopt certain morphologies. Mechanotransduction can then induce cells to start secreting collagen aligned with the surrounding extracellular matrix, thus further promoting cell organisation. 
    Understanding the interplay between  self-organising units and confinement can address both fundamental questions (e.g. is life a product of confinement or vice versa?) and help define design rules to steer self-organisation through confinement for applications. 
    
    \item \textbf{Design by confinement}: The final challenge is to identify and implement tangible design rules to predict emerging patterns from specific units (\emph{forward design}) or to optimise the units to obtain targeted spatio-temporal structures (\emph{inverse design}) under different realisations of confinement.  
    To enable optimal control, information needs to be encoded dynamically in both units and confining elements to generate complex, adaptable feedback mechanisms. 
    Biological systems are particularly good at encoding information (e.g.\ via DNA and RNA) and dynamically exploiting confinement to create function (e.g.\ by packing DNA in chromosomes within the cell nucleus or by assembling and disassembling functional compartments in cells, such as lysosomes or membrane-less organelles). There is broad scope for further developing man-made systems to mimic this biological complexity and harness emergence for technological applications, e.g. to develop programmable materials and smart devices for biomedicine \cite{grzybowski2017dynamic} or for crowd management \cite{feliciani2020systematic}.
    
\end{itemize}

\section*{Outlook}\label{outlook}

The above discussion highlights several avenues for future research, which, to be addressed, will require multiple conceptual and technological advancements. While methods and techniques are, of course, often system-specific, we expect the following open technical challenges to become relevant across scales and disciplines in the context of steering self-organisation through confinement.

First, we must develop improved tools to precisely characterise confinement, the interactions among the units, and the emergent structures. Experimentally, the nature and strength of confinement is not always easy to identify or quantify. 
This becomes particularly challenging for effective boundaries (e.g.\ for chemical gradients), moving boundaries due to their time dependence, and self-imposed forms of confinement that are intrinsically difficult to define and probe. Furthermore, the act of measurement might even alter the properties of the confining element itself, as already anticipated by Niels Bohr's complementarity principle for biology \cite{bohr1933light}.
Similarly, measuring the interactions among the units can pose a major challenge: in tissues for example, cell-cell interactions are influenced by a complex interplay of biochemical and mechanical signalling pathways and even by the constraints imposed by the surrounding medium \cite{Foolen2018}; in human crowds and animal groups, the interactions are influenced by psychological and cognitive factors that are difficult to quantify, especially given the intrinsic heterogeneity among individuals \cite{moussaid2009collective,von2016modelling}.  
Finally, it can be extremely challenging to dynamically probe the emerging self-organising structures experimentally from the outside: for example, due to partial or total opacity of the boundaries, real-time imaging with light microscopy can be problematic \emph{in vivo}, and the confinement itself can become a barrier to extract information about the self-organised system \cite{yoon2020deep}; in colloidal systems, interactions, while well-understood and measurable in bulk,  are less understood at liquid interfaces \cite{bradley2017janus}; \textit{in vivo} measurements can also be particularly difficult as the techniques used to probe the system can quickly become invasive enough to alter it (e.g. the phototoxicity and bleaching caused by fluorescence microscope imaging \cite{icha2017phototoxicity}).

Second, to develop a deeper understanding of how self-organisation can be steered through confinement, we must learn to identify and harness the key physical features both at a given scale and across scales. Notably, in the context of hierarchical confinement and reciprocity, one must first identify the relevant quantities that dictate the flow of information (e.g.\ pH, concentrations, mechanical forces, fluid velocity, chemical gradients, elasticity, etc.) and be able to measure these, before being able to understand the full cross-talk across scales. Here novel multi-scale and coarse-grained models will be particularly crucial,  
the development of which should occur in close synergy with experimental work to validate them. More generally, we must work towards improved experiments and models that are sufficiently simple and well-controlled to allow for scientific interpretation but which are also sufficiently detailed to capture the relevant phenomena observed under real-life conditions. This is especially imperative if we want to use these models to predict how different types of confinement and tailored units can steer self-organisation.

Lastly, to design by confinement, we must equip both the units and the boundaries with information-encoding and -processing degrees of freedom to enable adaptive feedback mechanisms.
Depending on the scales and systems of interest, the fabrication of information-encoding units and confinement can be achieved with techniques such as genetic engineering, nano- and microfabrication, 3D printing, or employing time-varying fields. A key technical challenge is the need for strong miniaturisation (as required by specific applications like precision medicine), which will limit the way we can design and control self-organising units and confinement at the smaller scales in future years. 
Finally, rapid progress in the field of machine learning is expected to guide the exploration of the enormous space of possibilities, both for new materials design (units and boundaries) and for the discovery of new self-organising structures in space and time \cite{wei2019machine}. 

In conclusion, steering self-organisation through confinement is a very active and rapidly evolving field of research, which is intrinsically multidisciplinary. To push the field forward, the scientific community working on self-organisation should increasingly take advantage of the cross-fertilisation of ideas that results from sharing hypotheses, theoretical approaches and experimental methods among experts from different fields and disciplines (e.g., between physical sciences and life sciences, between synthetic and natural systems, between small and large length scales). This cross-communication is {\it a priori} not easy, as it requires a common language and consensus on key open research questions and objectives. Certainly, the road ahead is still difficult and many steps need to be taken collectively to bring together the broader community and thus advance the field in a synergistic way. This perspective article provides a first step in this direction. We hope that it will serve as an impetus for the broader scientific community to join this collective effort and meet the exciting challenges that are faced across domains, length and time scales 
by the possibility of steering self-organisation through confinement. 

\section*{Acknowledgements}
All authors are grateful to the Lorentz Center for providing a venue for stimulating scientific discussions and to sponsor a workshop on the topic of ``Self-organisation under confinement'' along with the 4TU Federation, the J.M. Burgers Center for Fluid Dynamics and the MESA+ Institute for Nanotechnology at the University of Twente.
The authors are also grateful to Paolo Malgaretti, Federico Toschi, Twan Wilting and Jaap den Toonder for valuable feedback. 
N.A. acknowledges financial support from the Portuguese Foundation for Science and Technology (FCT) under Contracts no. PTDC/FIS-MAC/28146/2017 (LISBOA-01-0145-FEDER-028146), UIDB/00618/2020, and UIDP/00618/2020.
L.M.C.J. acknowledges financial support from the Netherlands Organisation for Scientific Research (NWO) through a START-UP, Physics Projectruimte, and Vidi grant.
I.C. was supported in part by a grant from by the Army Research Office (ARO W911NF-18-1-0032) and the Cornell Center for Materials Research (DMR-1719875).
O.D. acknowledges funding by the Agence Nationale pour la Recherche under Grant No ANR-18-CE33-0006 MSR.
M.D. acknowledges financial support from the European Research Council (Grant No. ERC-2019-ADV-H2020 884902 SoftML). 
W.M.D. acknowledges funding from a BBSRC New Investigator Grant (BB/R018383/1).
S.G. was supported by DARPA Young Faculty Award $\#$D19AP00046, and NSF IIS grant $\#$1955210.
H.G. acknowledges financial support from the Netherlands Organisation for Scientific Research (NWO) through Veni Grant No. 680-47-451. 
R.G. acknowledges support from the Max Planck School Matter to Life and the MaxSynBio Consortium, which are jointly funded by the Federal Ministry of Education and Research (BMBF) of Germany, and the Max Planck Society.
L.I. acknowledges funding from the Horizon Europe ERC Consolidator Grant ACTIVE$\_$ADAPTIVE (Grant No. 101001514).
G.H.K. acknowledges funding from the Netherlands Organisation for Scientific Research NWO through a VICI grant.
H.L. and N.V. acknowledge funding from the Deutsche Forschungsgemeinschaft (DFG) under grant numbers VO 1824/8-1 and LO 418/22-1. 
R.M. acknowledges funding from the Deutsche Forschungsgemeinschaft (DFG) under grant number ME 1535/12-1 and ME 1535/16-1. 
M.P. acknowledges funding from the Ram\'on y Cajal Program, grant no. RYC-2018-02534, and the Leverhulme Trust, grant no. RPG-2018-345.
A.Š. acknowledges financial support from the European Research Council (Grant No. ERC-2018-STG-H2020 802960 NEPA).
A.S. acknowledges funding from an ATTRACT Investigator Grant (No. A17/MS/11572821/MBRACE) from the Luxembourg National Research Fund. 
C.S. acknowledges funding from the French Agence Nationale pour la Recherche (ANR), grant ANR-14-CE090006 and ANR-12-BSV5001401, by the Fondation pour la Recherche Médicale (FRM), grant DEQ20120323737, and from the PIC3I of Institut Curie, France.
M.P. acknowledges funding from The Ram\'on y Cajal Program (Grant No. RYC-2018-02534) and from The Leverhulme Trust (Grant No. RPG-2018-345). 
I.T. acknowledges funding from the Spanish Ministerio de Ciencia e Innovación (MICINN) through the grant IED2019-000958-I. 
M.P. and I.T. also acknowledge funding from the Spanish Ministerio de Ciencia e Innovaci\'on (MICINN) through grant PID2019-104232GB-I00 and from the H2020 MSCA ITN PHYMOT (Grant agreement No 95591). 
I.Z. acknowledges funding from Project PID2020-114839GB-I00 MINECO/AEI/FEDER, UE. 
A.M. acknowledges funding from the European Research Council, Starting Grant No. 678573 NanoPacks.
G.V. acknowledges sponsorship for this work by the US Office of Naval Research Global (Award No. N62909-18-1-2170). 

\section*{Competing interests}
The authors have no competing interests.

\bibliography{sn-bibliography}


\end{document}